*Determination of the perturbations in the ionosphere produced by tsunamis through GNSS observations*

Domingo Centeno, Leonor Cui [(1) (2)]; Puente García, Víctor [(1) (2)]

[(1)] Área de Geodesia - Instituto Geográfico Nacional

General Ibáñez de Ibero, 3 – 28003 Madrid, España

[(2)] Facultad de Ciencias Matemáticas - Universidad Complutense de Madrid

Plaza de las Ciencias, 3 – 28040 Madrid, España

leonorcd@ucm.es, vpuente@mitma.es

## ABSTRACT

*During the propagation of a tsunami, gravity and sound waves can be produced, spreading from its source to the ionosphere's upper layers, thus generating perturbed electron densities in its E and F regions. These ionospheric disturbances can be studied in detail using measurements of the ionosphere's Total Electron Content (TEC), registered by permanent GNSS stations. In this contribution, the foundations of the VARION method (Variometric Approach for Real-time Ionosphere Observation) are described in order to obtain TEC's temporal variations with the aim of detecting such ionospheric disturbances. Moreover, the numerical results obtained after applying this method to real cases of tsunamis monitored by those satellites whose Ionospheric Pierce Points (IPPs) are closest to the tsunami source are presented. Lastly, based on these ionospheric perturbations reflected in the signals emitted by the satellites, a preliminary design is described for its potential integration into a Tsunami early Warning System (TWS) for the Iberian Peninsula.*

## 1. INTRODUCTION

In recent years, the demonstrated ability of GNSS to monitor a variety of events in an accurately, rapidly and cost-effectively way has allowed its use in countless different applications to grow considerably. Particularly, several studies have been carried out to analyze the behavior of the ionosphere in the event of natural hazards through the information that can be obtained, even in real time, from these GNSS observations. The ionosphere is the ionized part of the upper layers of the Earth's atmosphere, extending from 50 to 1000 kilometers above Earth's surface. This ionization is caused directly by the solar radiation activity, which modifies the properties of the ionosphere composition and produces disturbances on the ionospheric plasma densities. Depending on the degree of the ionization, different regions of the ionosphere can be considered with varying compositions at each height level. These regions are labelled as layers $D$, $E$, $F_1$ and $F_2$, where the $D$ region covers the upper part of the mesosphere, until the $F$ region reaching up to a part of the exosphere (Kelley, 2009).

In this article, we focus on the use of the ionosphere's properties for tsunami detection. According to UNESCO IOC NOAA International Tsunami Information Center[1], the majority of the tsunami's sources stem from earthquakes, though they can also be triggered by other phenomena such as earthquakes, landslides, volcanic eruptions, earthquakes, thunderstorms, deep convection events, space weather effects and a variety of anthropogenic events (explosions, rocket launches, etc.).

When a tsunami is created, it generates different types of acoustic and gravity waves that propagate upwards to the ionosphere. These waves can be grouped into two types, the acoustic-gravity waves (AGW), originated over the tsunami source zone, with a vertical velocity of 500-600 m/s, reaching the ionosphere 7-9 minutes after the main shock, and the internal gravity waves (IGW), caused due to the tsunami's propagation in the ocean, with a lower vertical velocity of 40-50 m/s (Astafyeva, 2019). The amplitude of these waves is amplified as they travel upward in the atmosphere due to the decrease in density as altitude rises. This makes the ionosphere a good medium for propagation and study, as the effect of these waves produces detectable disturbances in electron density specially in the $E$ and $F$ regions, at an altitude of between 300 and 350 km above sea level.

---

[1] http://itic.ioc-unesco.org/index.php

These changes in the ionosphere can be analyzed using measurements of the Total Electron Content (TEC) of the ionosphere — continuously collected by the operating GNSS ground-based receivers—, through a method based on a Variometric Approach for Real-time Ionosphere Observation, known by its acronym, VARION, as originally introduced in Savastano et al., (2017). Therefore, this study aims to analyze the numerical results obtained using this efficient and successful method in different potential tsunami hazard scenarios in order to outline the feasibility of integrating into a Tsunami early Warning System.

## 2. VARIOMETRIC APPROACH FOR REAL-TIME IONOSPHERE OBSERVATION METHOD

The Variometric Approach for Real-time Ionosphere Observation (VARION) method is a GNSS processing algorithm that focuses on real-time estimation of Slant Total Electron Content (STEC) variations. STEC is the measurement of the total number of free electrons in a unit cross column section along the ray path between the satellite and the receiver. This STEC measurement is represented in red in Figure 1.

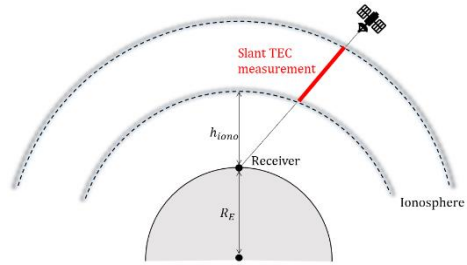

Figure 1. Slant TEC measurement between satellite-receiver ray

These estimations of STEC are affected by the transmitted signals as they travel through the ionosphere. Moreover, the estimates also depend on the local time (LT) hours of the day. Particularly, during the day, the ionosphere is more ionized than at night, reaching its highest degree of ionization between 12 LT and 16 LT hours. There is also a dependence on the receiver's latitude and longitude, as well as the solar and geomagnetic activity present in this layer.

Furthermore, the VARION method is based on the Variometric Approach for Displacements Analysis Stand-alone Engine, known by its acronym VADASE, which consists of an estimation of the ground velocities and displacements induced by several earthquakes in a real-time scenario (Benedetti et al., 2015).

### 2.1. MATHEMATICAL MODEL

The basis for the VARION method lies on single time differences of geometry-free combinations of GPS carrier-phase measurements, using a dual-frequency GPS receiver in a stand-alone operational mode. In this section, we describe the mathematical foundations of the VARION method, used for providing the time series of Slant TEC during the time in which a tsunami occurred with the aim of its detection.

Let $L_{iR}^S(t)$ be the standard raw carrier-phase observations, then in length units it is given by:

$$L_{iR}^S(t) = \rho_R^S(t) + c(\delta t_R(t) - \delta t^S(t)) + T_R^S(t) - I_{iR}^S(t) + \lambda_i N_{iR}^S(t) + p_R^S(t) + m_{iR}^S(t) + \varepsilon_R^S(t), \quad (1)$$

where the subscripts, $i$ and $R$, respectively correspond to the signal frequency and the receiver, and the superscript $S$ refers to the satellite. Furthermore, $\lambda$ is the carrier-phase wavelength, $\rho_R^S$ the geometric range, $c$ the speed of light, $\delta t_R$ and $\delta t^S$ are respectively the receiver and the satellite clock errors, $T_R^S$ and $I_R^S$ respectively refer to the tropospheric and the ionospheric delays along the satellite-receiver path, $N_{iR}^S$ is the ambiguity of carrier-phase, $p_R^S$ gathers the sum of other effects, such as variations on phase center of antenna, phase wind-up and the relativistic effects, and lastly, $m_{iR}^S$ and $\varepsilon_R^S$ correspond to the multipath effect and other residuals such as the noise, respectively.

Assuming that no cycle slips occur, the unknown ambiguity of carrier-phase remains constant between two consecutive epochs. Moreover, due to the dual frequency GPS observation, we consider the so-called geometry-free combination or $L_4$ combination. This combination cancels the geometric part of the measurement, leaving all the frequency-dependent effects. It is expressed as $L_4 = L_1 - L_2$, where $L_1$ and $L_2$ are respectively the GPS signals.

Therefore, differentiating (1) with respect to time between two consecutive epochs $(t, t+1)$, we obtain the geometry-free time single-difference observation equation:

$$L_{4R}^S(t+1) - L_{4R}^S(t) = \frac{f_1^2 - f_2^2}{f_2^2}\left(I_{1R}^S(t+1) - I_{1R}^S(t)\right), \quad (2)$$

wherein $f_1$ and $f_2$ are respectively the $L_1$ and $L_2$ carrier frequencies for GPS. Their values are derived from the fundamental frequency $f_0 = 10.23$ MHz as $f_1 = 154 f_0 = 1575.42$ MHz and $f_2 = 120 f_0 = 1227.60$ MHz.

Hence, considering the ionospheric refraction along the geometric range in (2), the Slant TEC variations between two consecutive epochs are:

$$\delta TEC(t) = \frac{f_1^2 f_2^2}{A(f_1^2 - f_2^2)}\left(L_{4R}^S(t+1) - L_{4R}^S(t)\right), \quad (3)$$

with $A$ being a constant derived from the plasma frequency given by Teunissen (2017):

$$A = \frac{1}{2}\frac{q_e^2}{4\pi^2 \epsilon_0 m_e} = 40.3081 \cdot 10^{16}\ \mathrm{m^3/s^2}, \quad (4)$$

where in turn $q_e$ is the electron charge, $\epsilon_0$ is the dielectric constant of vacuum and $m_e$ is the electron mass.

Afterwards, estimations of Slant TEC variations are computed by numerical integration of (3) over a time interval, $[t_0, t_f]$, during which the tsunami occurred. Thus, integration by the Trapezoidal Rule yields:

$$\Delta TEC(t_f, t_0) := \int_{t_0}^{t_f} \delta TEC(t)\, dt \approx \sum_{k=1}^{N} \frac{\delta TEC(t_{k-1}) + \delta TEC(t_k)}{2} \Delta t_k, \quad (5)$$

where $\Delta t_k = t_k - t_{k-1}$ is the time grid spacing. Without loss of generality, we can consider an equally spaced grid of size $\Delta t_k = h$, therefore having an error term of order $\mathcal{O}(h^2)$. These TEC measurements are expressed in TEC Units (TECU), where 1 TECU corresponds to $10^{16}$ electrons per square meter.

The VARION method may use the Klobuchar broadcast ionospheric model or even global maps, such as IONEX files, regarding the geolocation of the TEC measurements. Furthermore, it should be pointed out that this method may be implemented for all GNSS constellations, considering their observation codes, that is (L1C, L2W) for GPS, (L1C, L2C) or (L1P, L2P) for GLONASS, (L1C, L5Q) or (L1X, L5X) for Galileo and finally (L2I, L6I) or (L2I, L7I) for BeiDou.

Nevertheless, due to the ionospheric effects on the signals emitted between satellite and receiver, it is also crucial to analyze the geometry described by them during the time interval in which the tsunami occurs. The main objective consists of selecting those satellites that best monitor the tsunami.

### 2.2. CRITERIA FOR SATELLITE SELECTION

Most of the algorithms assume that the estimated TEC is mainly contributed by the $F_2$ region of the ionosphere, which justifies that the ionosphere can be modeled by a single-layer ionospheric layer, located at the height of the $F_2$ peak. This assumption allows us to attribute the estimated TEC to a specific point, termed as Ionospheric Pierce Point (IPP), defined as the point of intersection between the satellite-receiver line of sight and a thin ionospheric shell located at a fixed altitude, nominally taken as 350 km.

According to El-Gizawy (2003), the location of the IPP is computed using the ellipsoidal receiver coordinates in GRS80 geodetic reference system, $(\phi_r, \lambda_r)$, and the satellite ephemeris. This configuration, shown in Figure 2, allows us to obtain the Ionospheric Pierce Point (IPP) coordinates, $(\phi_{IPP}, \lambda_{IPP})$, given by:

$$\begin{cases} \phi_{IPP} = \phi_r + \psi \cos A, \\ \lambda_{IPP} = \lambda_r + \frac{\psi \sin A}{\cos \phi_{IPP}}, \end{cases} \quad (6)$$

where $A$ is the azimuth angle of the satellite and $\psi$ the offset angle, defined from the center of a supposedly spherical Earth, expressed as $\psi = E' - E$, with $E'$ and $E$ being the elevation angles at the IPP and the user's receiver respectively.

The elevation angle $E'$ can be expressed as:

$$E' = \arccos\left(\frac{R_E}{R_E + h_{iono}} \cos E\right), \tag{7}$$

being $R_E$ the mean radius of the spherical Earth and $h_{iono}$ the height of IPP, set as 350 km.

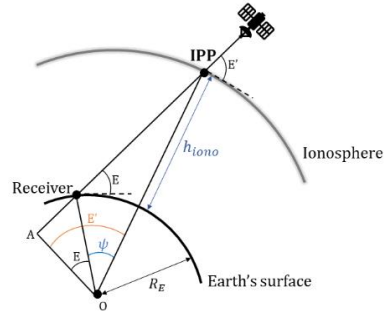

Figure 2. Satellite-receiver geometry

We seek to determine the satellite coordinates, expressed by its elevation and azimuth in (6), from the known receiver and satellite geodetic coordinates, expressed in the Earth-Centered Earth-Fixed (ECEF) coordinate system $(x, y, z)$. For this purpose, as defined in Subirana (2011), by means of a system transformation into the local East, North, Up (ENU) coordinate system —done through two rotation matrices—, we obtain the elevation and azimuth of the satellite in the local coordinate (ENU) system, given by:

$$\begin{cases} E = \arcsin(\vec{\rho} \cdot \vec{u}), \\ A = \arctan\left(\frac{\vec{\rho} \cdot \vec{e}}{\vec{\rho} \cdot \vec{n}}\right), \end{cases} \tag{8}$$

where $\vec{\rho}$ is the line of sight unit vector and finally, the unit vectors $\vec{e}, \vec{n}$ and $\vec{u}$ are defined as:

$$\begin{cases} \vec{e} = (-\sin\lambda_r, \cos\lambda_r, 0), \\ \vec{n} = (-\cos\lambda_r \sin\phi_r, -\sin\lambda_r \sin\phi_r, \cos\phi_r), \\ \vec{u} = (\cos\lambda_r \cos\phi_r, \sin\lambda_r \cos\phi_r, \sin\phi_r). \end{cases} \tag{9}$$

Depending on the moment magnitude scale, denoted by $Mw$, we define a threshold distance, $\kappa$, for those satellites whose IPPs are closest to the tsunami source coordinates, $(\phi_{ts}, \lambda_{ts})$. According to Kamogawa et al., (2016), this threshold distance is such that if $Mw \geq 8.4$, then $\kappa$ is set as 100 km while if $Mw < 8.4$, $\kappa$ takes 50 km. Hence, we selected those satellites whose IPPs coordinates lie within a circle of radius $\kappa$, centered in the tsunami source, this is:

$$(\phi_{IPP} - \phi_{ts})^2 + (\lambda_{IPP} - \lambda_{ts})^2 \leq \kappa^2. \tag{10}$$

## 3. REAL CASES OF TSUNAMIS

For the algorithm validation, we consider the ionospheric effects generated in the following real cases of tsunamis. Firstly, we consider the one that occurred in the Tohoku region of Japan, on the $11^{th}$ of March 2011 at 05:46:24 UTC. This undersea megathrust earthquake had a magnitude of 9.0 - 9.1 with its epicenter located at $\phi_{ts} = 38°.297\,N$ and $\lambda_{ts} = 142°.373\,S$, in the north-western Pacific Ocean, approximately 72 km east of the Oshika Peninsula of Tohoku, at a relatively shallow depth of 32 km. It was the most powerful earthquake ever recorded in Japan, and hence it is known as the *Great East Japan Earthquake*, lasting approximately six minutes and causing a consequential tsunami (Shibahara, 2011).

Besides, the Tonga tsunami in Oceania occurred on the $15^{th}$ of January 2022 at 04:14:45 UTC. It was caused by the eruption of a submarine volcano, Hunga Tonga–Hunga Ha'apai, in the Tonga archipelago located in the southern Pacific Ocean, whose coordinates are $\phi_{ts} = 20°.546\,S$ and $\lambda_{ts} = 175°.390\,W$. The eruption triggered several tsunamis, not only in Tonga, but also in Fiji, American Samoa, Vanuatu and along the Pacific coast, including damaging tsunamis in New Zealand, Japan, the United States, the Russian Far East, Chile and Peru. It was the largest volcanic eruption since the 1991 eruption of Mount Pinatubo, and the most powerful eruption since the 1883 eruption of Krakatoa (Denamiel, 2022).

In order to analyze the estimations of Slant TEC with the VARION algorithm, in the following section we describe the data used for each real case of tsunami.

## 4. DATASET

First of all, we selected the closest reference station receivers to the tsunami source from a map list of the International GNSS Service (IGS)[2]. Moreover, for the date on which the tsunami occurred, RINEX Observation files from those receivers are obtained from the Crustal Dynamics Data Information System (CDDIS)[3]. Finally, the orbit files, Standard Product 3 (SP3), followed by a series of position records and clock corrections for each satellite at each selected time epoch, are retrieved from Center for Orbit Determination in Europe (CODE)[4] product series. It should be noted that the satellite position is only needed to compute the position of the IPPs.

Given that the RINEX files are sampled every 30 seconds and the SP3 files every 15 minutes, we compute the satellite positions by means of a Lagrange interpolation, considering seventh-order interpolating polynomials, based on 4 points on each side of the emission time of signal, $T_{emission}$, which is computed as $T_{emission} = T_{reception} - \Delta t$, where $T_{reception}$ is the reception time of signal and $\Delta t$ the signal travel time, expressed as $\Delta t = \frac{Q}{c}$, being $Q$ the satellite-receiver pseudorange measure and $c$ the speed of light. Thus, for the validation of the algorithm, the numerical results for each real case of a tsunami are presented in the next section.

## 5. NUMERICAL RESULTS

To start with, for each case we show the closest stations chosen to the location where tsunami occurred and those that indeed have data in the respective RINEX and SP3 files. Through the RINEX file, we retrieved daily 30-second data for the Tohoku event and hourly 30-second data for Tonga event. Note that despite the fact that the Japanese region has a very dense GNSS network, for this date case, only GPS data were available (other constellations were not sampled). Besides, with regard to the SP3 files, final orbits with 15 min sample interval are considered for a better accuracy in the numerical results. This way, Figure 3 displays all the possible receivers for both cases, from which we chose the ones colored in red, leaving the green ones. Likewise, the Tohoku earthquake's epicenter and the location of the volcanic eruption for Tonga are represented with a blue pin symbol.

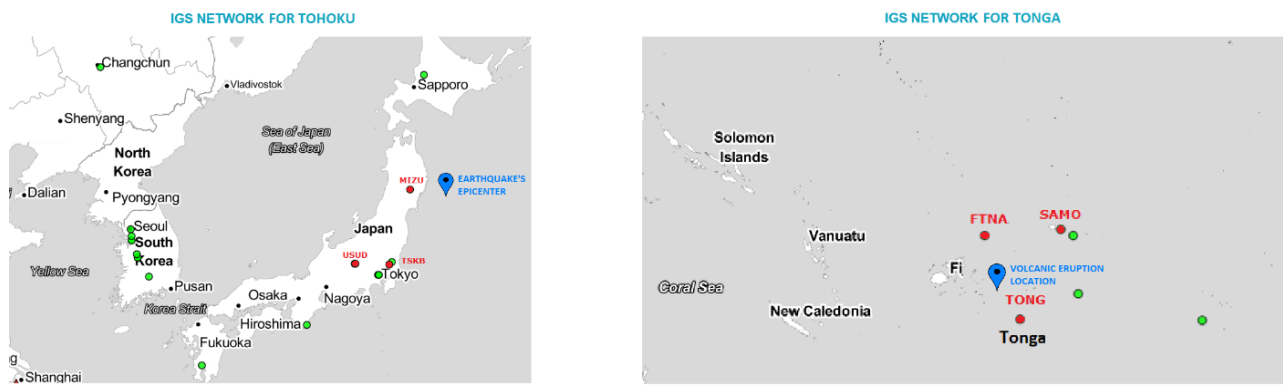

Figure 3. Map of possible IGS stations

As we can see, there are several possible stations for each event, however not all of them provided data for the respective tsunami dates, so we selected those in red being MIZU, USUD and TSKB stations for Japan and FTNA, TONG and SAMO for Tonga.

For each station, in Figure 4 all the GPS satellite tracks are shown, represented by their azimuth and elevation coordinates, expressed in the ENU coordinate system, and measured from the receivers chosen in a polar plot for each case of tsunami. In both cases, a mask of elevation of $10°$ has been considered in order to disregard satellites that appear close to the horizon as those at low angles present more atmospheric noise than satellites orbiting higher above the horizon, and they are subject to signal interference such as slip measurements or even multipath effect.

---

[2] https://igs.org/network/#station-map-list
[3] https://cddis.nasa.gov/archive/gnss/data/
[4] http://ftp.aiub.unibe.ch/CODE/

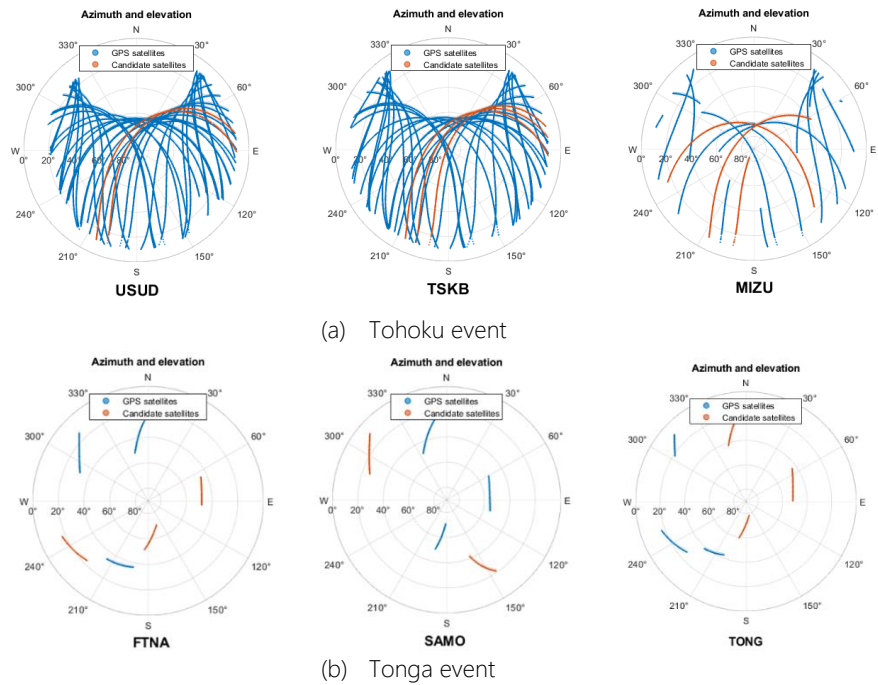

(a) Tohoku event

(b) Tonga event

Figure 4. Polar plot of GPS satellite's elevation and azimuth tracks

As it can be seen, owing to the threshold distance $\kappa$, dependent on the moment magnitude, of all possible satellites, colored in blue, we are left with those whose IPPs coordinates fulfill the equation of a circle established in (10). Therefore, the polar plot of the satellite paths at the Ionospheric Pierce Points in azimuth-elevation coordinates are colored in red. With regard to Figure 4, on the one hand, for the Tohoku event, we have 2,3 and 3 candidate satellites for each respective station USUD, TSKB and MIZU. On the other hand, we have 3,2 and 3 candidate satellites for FTNA, SAMO and TONG stations respectively. The number of satellites with their IPP coordinates closest to the tsunami source mainly depends on different aspects such as the GNSS data available for those specific dates; the fact that there may be no satellites in view from the station —since the emitted signal does not pass through the TEC depression in the ionosphere, called Tsunami Ionospheric Hole (TIH), which is produced by the waves after the tsunami occurrence over the area of its origin—; or even the aliasing effect that produces measurement errors in the signal due to an incorrectly adjusted sampling frequency. Note that the differences between Figure 4a and Figure 4b arise from the available datasets as we selected daily data for Tohoku's stations and hourly data for Tonga's stations.

In Figures 5 and 6 we plot the Slant TEC time series after applying the VARION method for both real cases of tsunami.

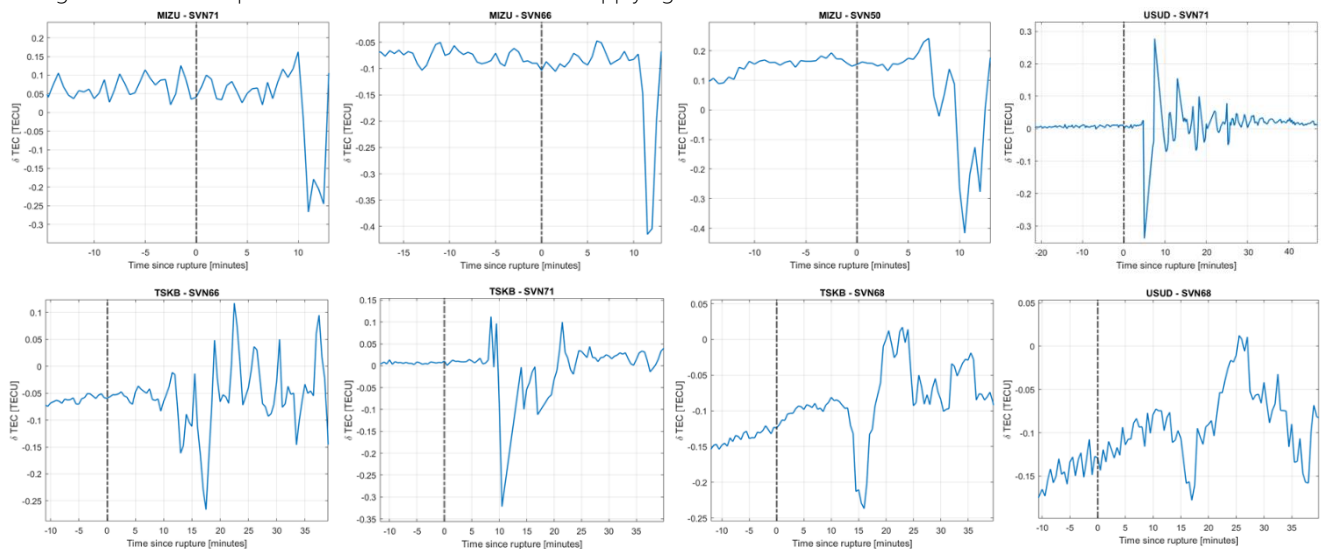

Figure 5. Slant TEC Time series monitored by GPS satellites for Tohoku region

10th Spanish-Portuguese Assembly of Geodesy and Geophysics - Toledo, November 2022    6

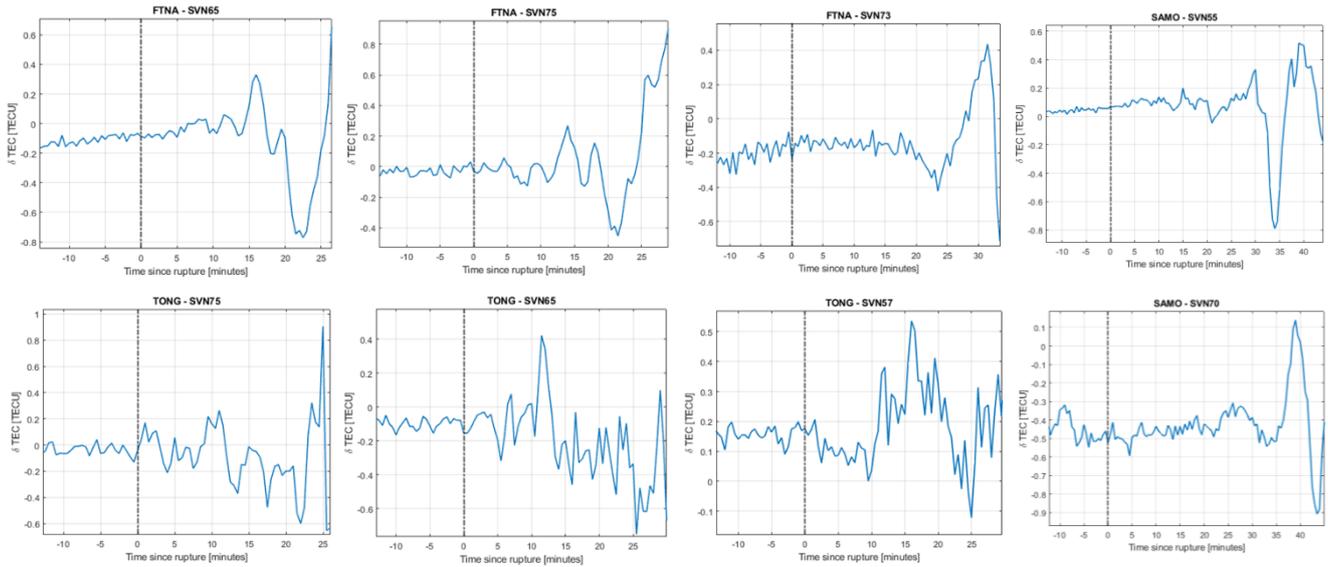

Figure 6. Slant TEC Time series monitored by GPS satellites for Tonga

As it is shown, the dashed line indicates the moment when the event source was triggered, so all the time series are computed from the time of occurrence, spanning from 10 minutes before the rupture to as late as 40 minutes after it, depending on the case. Note that, for each station we identify the candidate satellites with their SVN number instead of the satellite's PRN for uniqueness. In addition, leap seconds are added for more accurate estimations, considering 15 and 18 leap seconds for Tohoku and Tonga events respectively.

These numerical results show the Slant TEC behavior. At the first stages, the time series from Figures 5 and 6 are statistically stationary, verifying that their mean and variance are both constant over this time period. However, minutes after the main shock, the Slant TEC values begin to oscillate with no clear trend, reflecting the perturbations on the ionosphere layer caused by these tsunamis. As it can be seen, related to the necessary detection time from the origin of the event to detect ionospheric anomalies, we have the following. For the Tohoku event, all the selected stations monitor these disturbances in less than 15-20 minutes, where the amplitude of the Slant TEC disturbances varies in absolute value in a range of 0.4 TECU. On the other hand, for the Tonga event, the chosen stations take a few minutes more for monitoring the perturbations, possibly because of the satellite arrangement. For this case, significant amplitude of Slant TEC disturbances about $\pm 1.0$ TECU are found minutes after the volcanic eruption, higher than normal.

Lastly, for analyzing the results, in Figure 7 we plot only the track of candidate satellites whose IPPs coordinates are closest to the tsunami source. These IPPs coordinates are colored in red, while the source and receiver coordinates are respectively colored in blue and green.

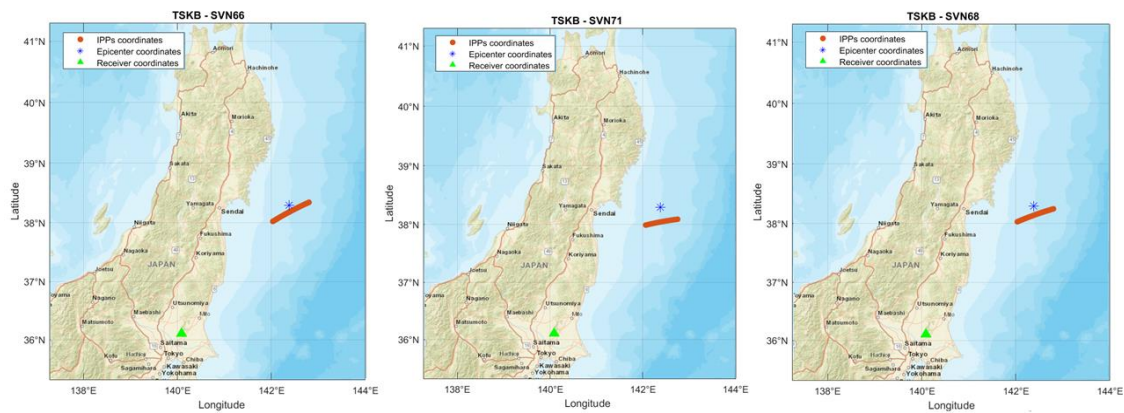

(a) Tohoku event

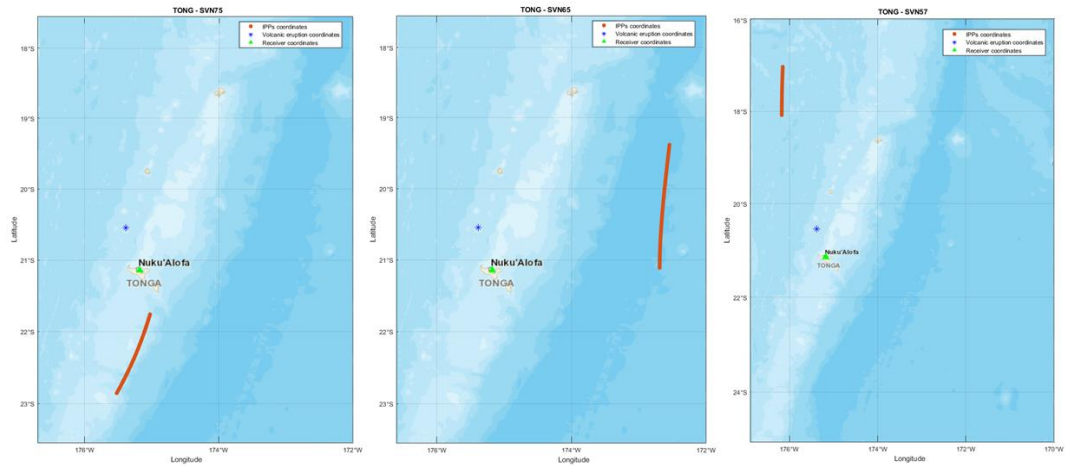

(b) Tonga event

Figure 7. Candidate satellite tracks at the Ionospheric Pierce Points

Through these numerical results, the VARION algorithm has demonstrated to be capable of detecting ionospheric perturbations generated by tsunami-driven gravity waves and can be considered as a novel contribution in future operations of Tsunami early Warning Systems.

## 6. TSUNAMI WARNING SYSTEMS

Tsunami Warning Systems (TWS) are based on detecting tsunamis in advance, issuing warnings on those potentially destructive tsunamis and effectively preventing tsunami disasters. This requires tsunami forecasts to be made in real time so that affected locations receive evacuation warnings in time to avoid casualties and large damages to civilian properties. Hence, for their prevention, they verify the tsunami occurrence before its waves arrive in the coasts by the features that the GNSS measurements should have in order to alert which locations would be affected, the arrival time, the wave height estimation on each one of those locations as well as the confirmation of whether the tsunami has actually been generated or not.

As previously mentioned, the main sources of tsunamis are due to earthquakes, therefore the current systems are mainly based on the use of seismometers, rapidly detecting the location and size of the potentially destructive earthquakes by recording the motion of the ground. Thus, considering the seismic parameters obtained, these systems use different tools to provide risk assessments, such as numerical simulations, decision matrices or even tsunami scenario databases processed by central processing facilities, to assess whether the detected earthquake may trigger a tsunami. For instance, numerical tsunami simulations are mainly based on the estimation of the tsunami's wave properties or even the estimation of the initial tsunami height using the long-period seismic wave records, where the vertical displacement of the rupture area is required to be calculated at the same time (Blewitt et al., 2009). For this purpose, early warning systems commonly use the MOST (Method of Splitting Tsunami) numerical model from the National Oceanic and Atmospheric Administration's (NOAA) Tsunami Warning System, which uses a finite difference scheme for the characteristic form of the shallow water wave equations, thus providing an estimation of the tsunami wave height. Similarly, DART/WP-GITM software uses the ionospheric perturbations in order to estimate the tsunami wave properties through a tsunami parameter inversion model (Meng et al., 2015).

Owing to the tens of additional minutes that are required for the accurate calculation of the rupture area, it is difficult to issue a warning of the precise height of the next near-field tsunami early enough for residents in coastal areas to evacuate. To overcome this problem, offshore tsunamis are monitored using positioning system (GPS) buoys, as they issue faster confirmation, linked to ocean bottom pressure sensors and GNSS satellites in order to provide high accuracy in the measurements of tidal level. For instance, the Deep-ocean Assessment and Reporting of Tsunamis (DART) buoys are located 300 km away from coasts and constantly monitor the changes in the water columns (Meinig, 2005). Nonetheless, the acquisition, installation and maintenance costs of a sufficiently dense network for such a monitoring system would require a high budget and manpower. Therefore, less expensive practical Tsunami early Warning Systems are urgently needed. Moreover, the use of tide gauges, located at the coastlines, monitoring continuously the changes in the water height or even if the use of GNSS satellites as well as satellite altimeters measuring the ocean surface height, may help in the detection of tsunami.

However, there are several uncertainties in tsunami's prediction such as inaccurate knowledge of the marine active faults, no direct match between earthquake's size and tsunami's severity, the estimation of the hypocentral location and its magnitude, or even if the lack of information on the current sources of the earthquake, i.e., rupture extent, fault geometry, direction of slip, etc. In addition, other anomalies can be encountered which affect the ionosphere unrelated to tsunami generation. They are mainly produced by external origins such as ionization difference due to changes caused by solar radiation activity, geomagnetic storms or even to a lesser degree, strong meteorological process may affect this ionosphere layer.

For this purpose, the Tsunami early Warning System described in this article, based on using GNSS satellites for ionosphere monitoring, overcomes most of these limitations by means of the potential integration of the VARION method into these warning systems. In particular, the possibility of incorporating this method into the Spanish Tsunami Warning System is discussed in the following section.

## 7. SPANISH TSUNAMI WARNING SYSTEM

Historically, the Iberian Peninsula has suffered several tsunamis in the past, notably in the Atlantic area of the southwest of the Iberian Peninsula as can be seen in Figure 8.

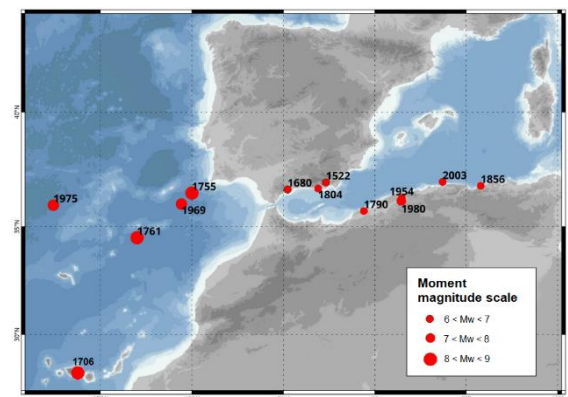

Figure 8. Historical tsunamis in the Iberian Peninsula. Adapted from Cantavella (2021).

The largest known historical tsunami affected the Iberian Peninsula's coast in 1755, after the devastating effects caused by an earthquake of magnitude 8.5 in Lisbon, simply known as the *Lisbon earthquake*, ravaging several Portuguese and Spanish cities, and also causing thousands of deaths. Its hypocenter was located offshore, southwest of the Iberian Peninsula, in the Azores-Gibraltar fault zone, from where waves of up to 10-15 meters on the coast of the Gulf of Cadiz were produced (Oliveira, 2008). On the other hand, moderate tsunamis have occurred on the Mediterranean coast, located near the northern coast of Algeria and mainly in the Alboran Sea and in the faults in the Gulf of Cadiz, where the Spanish tsunami threat is maximum likelihood (Aniel-Quiroga, 2017).

Considering this background, the Spanish Tsunami Warning System (STWS) was implemented by the National Geographic Institute of Spain (IGN)[5], being operational since November 2015. It consists of an automatic procedure that evaluates whether a tsunami may affect the Spanish coasts. In this way, in case of a tsunami threat, an alert message is sent to the National Civil Protection authorities with an estimate of the degree of such hazard at different locations along the coasts, the estimated severity of the tsunami and the time of arrival for those places. In addition, several Spanish stations collect multi-frequency data from GPS satellites, while others track multi-GNSS data from other constellations (GLONASS, Galileo and BeiDou), allowing us to better analyze the ionosphere. Moreover, with the help of the ionosonde, situated in "El Arenosillo", the analysis of the ionogram provided, and the geomagnetic observatories, located in San Pablo, San Fernando and Güímar, will allow us to discern the sources of ionospheric disturbances not related to tsunamis, such as geomagnetic storms.

For these reasons, we propose to implement and integrate this complementary novel approach, the VARION method, into the Spanish Tsunami Warning System (STWS) in order to detect the occurrence of a tsunami before its waves reach the coasts of the Iberian Peninsula.

---

[5] http://www.ign.es

It should be noted that to incorporate the ionospheric anomalies generated by tsunamigenic earthquakes into the Tsunami Warning Systems, their magnitude ranges that produce disturbances in the ionosphere have been analyzed. In Le et al., (2011), the authors establish that earthquakes with a Mw magnitude lower than 6 produce less obvious ionospheric disturbances in TEC, causing small perturbations difficult to remark and thus generating confusion with other events, such as geomagnetic storms, severe meteorological processes, etc.

Numerous studies are currently underway to improve these Tsunami early Warning Systems, including the development of new implementations. According to Martire et al., (2023), the GUARDIAN (GNSS Upper Atmospheric Real-time Disaster Information and Alert Network) system, is an ionospheric monitoring software in near-real time (NRT) through multi-GNSS TEC measurements in order to augment the early natural disaster warnings. Related to these initiatives, close international collaborations are maintained with the aim of disaster risk reduction. For instance, there is a special commission within the International Union of Geodesy and Geophysics called Geophysical Risk and Sustainability (IUGG GeoRisk)[6] with the aim of researching on geophysical hazards and the disaster preparedness. Secondly, the International Association of Geodesy (IAG) includes the GGOS Geohazards Focus Group[7] for enhancing the GNSS-based Tsunami Early Warning Systems (GTEWS) as well as many member countries from United Nations collaborate under the agreement on the Sendai Framework for Disaster Risk Reduction with the goal of upgrading the disaster preparedness.

## 8. CONCLUSIONS

In this paper we analyzed the perturbations in the ionosphere caused by a tsunami through Total Electron Content (TEC) measurements, steadily registered by permanent GNSS stations. For this tsunami detection, we implemented a real-time approach termed as VARION (Variometric Approach for Real-time Ionosphere Observation) method, characterized by its independence of cycle slips and a very low computational time. This method provides estimations of the Slant TEC (STEC) time series during the time interval in which the tsunami occurred, that are monitored by those GNSS satellites whose Ionospheric Pierce Point (IPP) coordinates were closest to the source.

Its demonstration has been performed on two real cases of tsunami, one of them recorded in the Tohoku region, caused by an earthquake; and the other one occurring in Tonga due to a volcanic eruption. The analysis of the numerical results plotted in Figure 5 and 6 proved the expected variations in the Slant TEC time series since the moment of rupture, thus reflecting the consequent disturbances on the ionosphere caused by these tsunamis.

This approach has a low cost of implementation as well as it reduces the confirmation time of events, since the method takes advantage of the existing infrastructures such as the GNSS derived data and the stations available in the area for monitoring the ionospheric disturbances. Thus, combining the GNSS satellites integration and operation capabilities with its GNSS receiver development, we have presented this novel method to be integrated into a Tsunami Warning System (TWS) and more specifically into the Spanish Tsunami Warning System (STWS) as an auxiliary technique that complements the existing methodologies in the verification of potential tsunami hazards for the Iberian Peninsula.

For future lines of work, the automatization of VARION algorithm by using GNSS broadcast real-time data combined with real-time data from different sources, such as tide gauges, seismometers, buoys or even GNSS receivers are considered. The large amount of available data at no additional cost, is obtained through the multiple satellite constellations orbiting the Earth and the number of ground-based GNSS receivers placed worldwide. Thus, once an event is detected at a specific location, such a system would begin processing the Slant TEC through real-time data using multiple available stations located near the tsunami source in search of disturbed ionospheric signals minutes after the atmospheric wave reaches the ionosphere. Its continuous analysis also considers the stations that have available data in their RINEX files; the orbit positions and clock offsets of GPS satellites; and finally the selection of station-satellite pairs that verify that their IPP coordinates are within the previously established threshold distances from the source. The above mentioned reasons make GNSS-based real-time monitoring of the ionosphere an interesting enhancement for natural hazard early warning systems.

---

[6] https://iugg.org/associations-commissions/commissions/
[7] https://ggos.org/about/org/fa/geohazards/